# Orthotropic metamaterials with freely tailorable elastic constants

Jiang Ke

(School of Civil Engineering and Architecture, Shaanxi University of Technology, Hanzhong 723001, China)

**Abstract**   An orthotropic metamaterial is composed of elements arrayed periodically in space. The element includes two cuboid structures. The first structure is the basic structure of the element, and the second structure is the transformation of the first structure of the element. The first structure of the element is a cuboid structure composed of 24 bars connected by 8 nodes, and the second structure of the element is a cuboid structure composed of 36 bars connected by 14 nodes. This metamaterial has 6 independent elastic constants, so there is a large degree of freedom in material design. Using a simple universal design method, a metamaterial with tailored elastic constants can be designed. Therefore, it has great application value in the fields of mechanical metamaterials, elastic wave metamaterials, acoustic metamaterials, and seismic metamaterials, and has also laid the foundation for realizing the dream of controlling elastic waves, acoustic waves and vibrations.

**Key words**   metamaterial, elastic wave, acoustic wave, orthotropic, negative Poisson's ratio, zero Poisson's ratio

**Introduction**

The propagation control of elastic waves has great application value in aerospace, submarine, mechanical engineering, civil engineering and other fields. According to the theory of elastodynamics, the elastic constants of a material are the major factors affecting the propagation of elastic waves. Therefore, an important method to control the propagation of elastic waves is to design elastic wave metamaterials. In addition, mechanical metamaterials with negative Poisson's ratio or zero Poisson's ratio, acoustic metamaterials controlling acoustic waves and seismic metamaterials controlling seismic waves, all have great application value and are also emerging research hotspots. So far, the metamaterials studied are mainly isotropic metamaterials and metamaterials with anisotropic mass density, which are difficult to design. Obviously, orthotropic metamaterials can better control elastic waves, acoustic waves and vibrations, but the design of such metamaterials is more difficult. An orthotropic material has 12 elastic constants, including 6 Poisson's ratios, 3 elastic moduli and 3 shear moduli, but only 9 independent elastic constants. If the elastic constants of a material can be designed relatively freely, it can be used to design mechanical metamaterials, elastic wave metamaterials, acoustic metamaterials and seismic metamaterials. However, it is very difficult to design metamaterials with tailored elastic constants. At present, there is no simple and effective universal design method [1-15].

In 2012, based on the generalized Hooke's law and the superposition principle, Jiang Ke proposed a space truss element model for calculating orthotropic elastic problems, i.e., calculating the stresses, strains and displacements of each point in the solid under load. The model is applied to materials with 9 elastic constants satisfying 3 conditions, i.e., the material has only 6 independent elastic constants [16]. Based on the space truss element model, an orthotropic metamaterial and its design method are proposed in this paper.

**1 An orthotropic metamaterial and its design method**

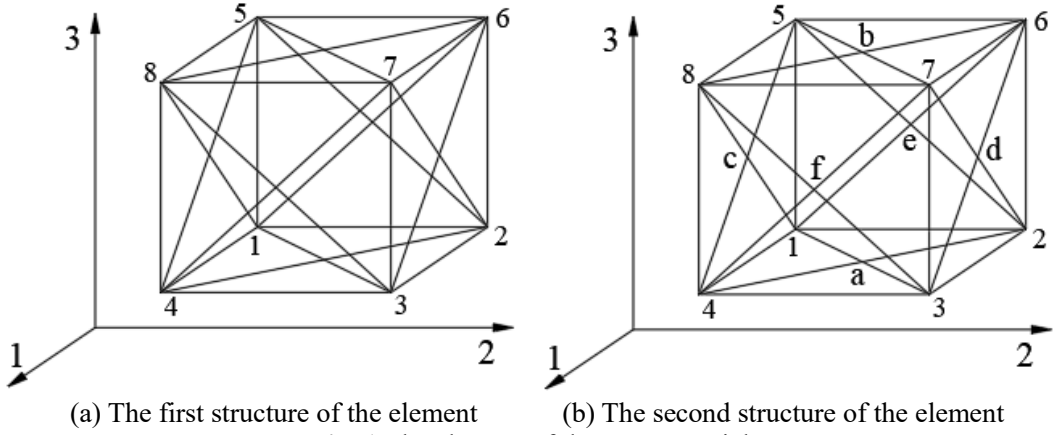

(a) The first structure of the element  (b) The second structure of the element
**Fig. 1** The element of the metamaterial

An orthotropic metamaterial is composed of elements arrayed periodically in space. The element includes two cuboid structures. The first structure is the basic structure of the element, and the second structure is the transformation of the first structure of the element, as shown in Fig. 1. The 1-, 2-, and 3-directions of the coordinate system in Fig. 1 are the three principal material directions.

The first structure of the element is a cuboid structure composed of 24 bars connected by 8 nodes, as shown in Fig. 1(a); where, there are 4 bars parallel to each principal material direction, for 3 principal material directions, then there are 12 parallel bars in total; there are 2 oblique bars on each face of the cuboid structure, for 6 faces, then there are 12 oblique bars in total; the 8 nodes used to connect the bars are 1, 2, 3, 4, 5, 6, 7 and 8.

The second structure of the element is a cuboid structure composed of 36 bars connected by 14 nodes, as shown in Fig. 1(b), compared with the first structure of the element, the difference is that there are 4 oblique bars on each face of the cuboid structure, for 6 faces, then there are 24 oblique bars in total; 6 nodes are added, i.e. a, b, c, d, e and f.

The dimensions of the element in the 1-, 2-, and 3-directions are:

$l_1=l_{14}=l_{23}=l_{67}=l_{58}$,  $l_2=l_{12}=l_{56}=l_{87}=l_{43}$,  $l_3=l_{15}=l_{48}=l_{37}=l_{26}$.

where, $l$ is the length of the bar, for example, $l_{26}$ is the length of the bar 26, 2 and 6 is its two nodes.

According to the value of the axial stiffness, the axial stiffnesses of all bars in the element are divided into 6 types, as follows:

$K_1=K_{14}=K_{23}=K_{67}=K_{58}$,  $K_2=K_{12}=K_{56}=K_{87}=K_{43}$,  $K_3=K_{15}=K_{48}=K_{37}=K_{26}$,

$K_4=K_{13}=K_{24}=K_{57}=K_{68}=K_{1a}=K_{2a}=K_{3a}=K_{4a}=K_{5b}=K_{6b}=K_{7b}=K_{8b}$,

$K_5=K_{18}=K_{54}=K_{27}=K_{63}=K_{1c}=K_{5c}=K_{8c}=K_{4c}=K_{2d}=K_{6d}=K_{7d}=K_{3d}$,

$K_6=K_{16}=K_{25}=K_{47}=K_{38}=K_{1e}=K_{2e}=K_{6e}=K_{5e}=K_{4f}=K_{3f}=K_{7f}=K_{8f}$.

where, $K$ is the axial stiffness of the bar, for example, $K_{8f}$ is the axial stiffness of the bar 8f, 8 and f is its two nodes.

$K_1$ is the axial stiffness of one parallel bar parallel to the 1-direction,

$K_2$ is the axial stiffness of one parallel bar parallel to the 2-direction,

$K_3$ is the axial stiffness of one parallel bar parallel to the 3-direction,

$K_4$ is the axial stiffness of one oblique bar parallel to the 1-2 plane,

$K_5$ is the axial stiffness of one oblique bar parallel to the 3-1 plane,

$K_6$ is the axial stiffness of one oblique bar parallel to the 2-3 plane.

Any bar in the element can be replaced by a spring or a variable cross-section bar according to the principle of keeping the length and stiffness coefficient of the bar unchanged. The stiffness coefficient of a bar is the axial force when the bar produces a unit axial displacement, which is equal to the ratio of the axial stiffness of the bar to the length of the bar.

The design method of the metamaterial generally uses the following steps.

Step 1: The design goal of the metamaterial is to tailor *n* independent elastic constants in the 9 elastic constants, where, $1 \leq n \leq 6$.

According to the design goal of the metamaterial, the design values of 9 elastic constants of the metamaterial are determined. The tailored n independent elastic constants are known, and then the (6-*n*) independent elastic constants are freely selected, and the other 3 elastic constants are determined by the following formula (1).

$$G_{12} = \frac{(v_{12} + v_{13}v_{23}E_2/E_3)E_1}{(1 - v_{12}^2 E_1/E_2 - v_{13}^2 E_1/E_3 - v_{23}^2 E_2/E_3 - 2v_{12}v_{13}v_{23}E_1/E_3)} ,$$

$$G_{31} = \frac{(v_{13} + v_{12}v_{23})E_1}{(1 - v_{12}^2 E_1/E_2 - v_{13}^2 E_1/E_3 - v_{23}^2 E_2/E_3 - 2v_{12}v_{13}v_{23}E_1/E_3)} ,$$

$$G_{23} = \frac{(v_{23}E_2/E_1 + v_{12}v_{13})E_1}{(1 - v_{12}^2 E_1/E_2 - v_{13}^2 E_1/E_3 - v_{23}^2 E_2/E_3 - 2v_{12}v_{13}v_{23}E_1/E_3)} \quad (1)$$

where, $E_1$, $E_2$, $E_3$ = elastic moduli in the 1-, 2-, and 3-directions,

$v_{ij}$ = Poisson's ratio, i.e., the negative of the transverse strain in the i-direction over the strain in the j-direction when stress is applied in the j-direction, i.e., $v_{ij} = -\varepsilon_i/\varepsilon_j$ for $\sigma = \sigma_j$ and all other stresses are zero,

$G_{23}$, $G_{31}$, $G_{12}$ = shear moduli in the 2-3, 3-1, and 1-2 planes.

Step 2: Determine whether the 9 elastic constants of the metamaterial are within the allowable range of orthotropic elasticity, and the allowable range is determined by Formula (2).

$E_1 > 0$, $E_2 > 0$, $E_3 > 0$,

$|v_{12}| < (E_2/E_1)^{1/2}$, $|v_{23}| < (E_3/E_2)^{1/2}$, $|v_{13}| < (E_3/E_1)^{1/2}$,

$G_{12} > 0$, $G_{31} > 0$, $G_{23} > 0$,

$1 - v_{12}^2 E_1/E_2 - v_{13}^2 E_1/E_3 - v_{23}^2 E_2/E_3 - 2v_{12}v_{13}v_{23}E_1/E_3 > 0$,

$$1 - v_{12}^2 E_1/E_2 > 0, \quad 1 - v_{13}^2 E_1/E_3 > 0, \quad 1 - v_{23}^2 E_2/E_3 > 0 \quad (2)$$

If all 9 elastic constants of the metamaterial are within the allowed range, proceed to the next step, otherwise, reselect the elastic constants and repeat the previous step and this step. It can also be checked after step 3.

Step 3: Select the dimensions $l_1$, $l_2$, $l_3$ in the three directions of the element, and then calculate the axial stiffness of each bar in the element by Formula (3).

$$K_1 = \frac{l_2^2 l_3^2 (1 - v_{23}^2 E_2/E_3) - l_1^2 l_2^2 (v_{13} + v_{12}v_{23}) - l_1^2 l_3^2 (v_{12} + v_{13}v_{23}E_2/E_3)}{4 l_2 l_3 (1 - v_{12}^2 E_1/E_2 - v_{13}^2 E_1/E_3 - v_{23}^2 E_2/E_3 - 2v_{12}v_{13}v_{23}E_1/E_3)/E_1} ,$$

$$K_2 = \frac{l_1^2 l_3^2 (1 - v_{13}^2 E_1/E_3)E_2/E_1 - l_1^2 l_2^2 (v_{23}E_2/E_1 + v_{12}v_{13}) - l_2^2 l_3^2 (v_{12} + v_{13}v_{23}E_2/E_3)}{4 l_1 l_3 (1 - v_{12}^2 E_1/E_2 - v_{13}^2 E_1/E_3 - v_{23}^2 E_2/E_3 - 2v_{12}v_{13}v_{23}E_1/E_3)/E_1} ,$$

$$K_3 = \frac{l_1^2 l_2^2 (1 - v_{12}^2 E_1 / E_2) E_3 / E_1 - l_1^2 l_3^2 (v_{23} E_2 / E_1 + v_{12} v_{13}) - l_2^2 l_3^2 (v_{13} + v_{12} v_{23})}{4 l_1 l_2 (1 - v_{12}^2 E_1 / E_2 - v_{13}^2 E_1 / E_3 - v_{23}^2 E_2 / E_3 - 2 v_{12} v_{13} v_{23} E_1 / E_3) / E_1} \;,$$

$$K_4 = \frac{l_3 (l_1^2 + l_2^2)^{1.5} (v_{12} + v_{13} v_{23} E_2 / E_3) E_1}{4 l_1 l_2 (1 - v_{12}^2 E_1 / E_2 - v_{13}^2 E_1 / E_3 - v_{23}^2 E_2 / E_3 - 2 v_{12} v_{13} v_{23} E_1 / E_3)} \;,$$

$$K_5 = \frac{l_2 (l_1^2 + l_3^2)^{1.5} (v_{13} + v_{12} v_{23}) E_1}{4 l_1 l_3 (1 - v_{12}^2 E_1 / E_2 - v_{13}^2 E_1 / E_3 - v_{23}^2 E_2 / E_3 - 2 v_{12} v_{13} v_{23} E_1 / E_3)} \;,$$

$$K_6 = \frac{l_1 (l_2^2 + l_3^2)^{1.5} (v_{23} E_2 / E_1 + v_{12} v_{13}) E_1}{4 l_2 l_3 (1 - v_{12}^2 E_1 / E_2 - v_{13}^2 E_1 / E_3 - v_{23}^2 E_2 / E_3 - 2 v_{12} v_{13} v_{23} E_1 / E_3)} \quad (3)$$

For any bar in the element, select the elastic modulus of this bar material, and then calculate the cross-sectional area of this bar, i.e., the cross-sectional area of this bar = the axial stiffness of this bar / the elastic modulus of this bar material.

Check the calculated cross-sectional area of each bar, if there is a negative cross-sectional area, adjust the dimensions of the element, and then recalculate according to this step, if there is still a negative cross-sectional area, modify the design values of the elastic constants of the metamaterial and redesign from the first step, if there is no negative cross-sectional area of each bar, select the cross-sectional shape of this bar and determine the cross-sectional size of this bar, so that the cross-sectional size of all bars in the element can be obtained, where, if the cross-sectional area of a bar is 0, it means that this bar is not included in the element. Then, select the first structure or the second structure from the two structures of the element; finally, according to the dimensions of the element, determine whether the cross-section of each bar is too large in the space range of one element, resulting in a crowded space and unable to connect each bar to form one element. If there is such a problem, increase the elastic modulus of each bar material, and then re-determine the cross-sectional area and cross-sectional size of each bar.

The elements are arrayed periodically in space, and the adjacent elements share their adjacent nodes, thus forming the metamaterial. If in order to reduce the number of bars in the metamaterial, the adjacent bars of the adjacent elements are merged into one bar by summing their cross-sectional areas.

It is possible that there is a deviation between the actual value of the elastic constant of this metamaterial and its design value. In order to evaluate the accuracy of the design, the finite element method can be used, i.e., take one element and calculate it under uniaxial tension (or uniaxial compression) in the three principal material directions respectively, and then the 3 elastic moduli and 3 Poisson's ratios of the metamaterial can be determined by the definitions of elastic modulus and Poisson's ratio, calculate under pure shear in 1-2 plane, 2-3 plane and 3-1 plane respectively, and then the 3 shear moduli of the metamaterial can be determined by the definition of shear modulus.

Various connection modes can be used between the bars in the element, but the actual connection mode is generally between the hinged connection and the rigid connection. When calculating, it can be simplified according to the specific situation. If the actual connection mode is close to the hinged connection, then it is considered as hinged connection, otherwise, it is considered as rigid connection. In order to design the metamaterial with small shear modulus and remove the oblique bars in the element, the shear modulus in step 1 is set equal to 0, but the connection mode between the bars in the element is considered as rigid connection, the actual shear modulus of the metamaterial is greater than

0, which still satisfies the requirements of elasticity in step 2.

## 2 Examples of metamaterial design

In order to illustrate the design method of metamaterials, 6 examples with negative Poisson's ratio or zero Poisson's ratio are given below, where, Example 1 is described in detail, and the other examples are only briefly described.

### 2.1 Example 1

The design goal of this orthotropic metamaterial is to tailor one Poisson's ratio $v_{12}$= -0.5 in the 9 elastic constants.

Step 1: Based on the design goal of this metamaterial, determine the design values of the 9 elastic constants of this metamaterial, known $v_{12}$= -0.5, and then freely select 5 independent elastic constants, $v_{23}$=0.85, $v_{13}$=0.75, $E_1$=480 N/mm$^2$, $E_2$=400 N/mm$^2$, $E_3$=500 N/mm$^2$, and the other 3 elastic constants are obtained from Formula (1), i.e., $G_{12}$=24.7423 N/mm$^2$, $G_{31}$=804.124 N/mm$^2$, $G_{23}$=824.742 N/mm$^2$.

Step 2: From the formula (2), $E_1$=480 N/mm$^2$ >0, $E_2$=400 N/mm$^2$ >0, $E_3$=500 N/mm$^2$ >0, $|v_{12}|$=0.5<$(E_2/E_1)^{1/2}$=0.91, $|v_{23}|$=0.85<$(E_3/E_2)^{1/2}$=1.12, $|v_{13}|$=0.75<$(E_3/E_1)^{1/2}$=1.02, $G_{12}$=24.7423 N/mm$^2$ >0, $G_{31}$=804.124 N/mm$^2$ >0, $G_{23}$=824.742 N/mm$^2$ >0,

$$1 - v_{12}^2 E_1 / E_2 - v_{13}^2 E_1 / E_3 - v_{23}^2 E_2 / E_3 - 2 v_{12} v_{13} v_{23} E_1 / E_3 = 0.194 > 0,$$

$$1 - v_{12}^2 E_1 / E_2 = 0.7 > 0, \quad 1 - v_{13}^2 E_1 / E_3 = 0.460 > 0, \quad 1 - v_{23}^2 E_2 / E_3 = 0.422 > 0.$$

Therefore, the 9 elastic constants of this metamaterial are all within the allowable range of orthotropic elasticity.

Step 3: Select the dimensions in the three directions of the element, $l_1$=10mm, $l_2$=10mm, $l_3$=10mm, then by Formula (3), the axial stiffnesses of the bars in the element are:
$K_1$=5381.44 N, $K_2$=2474.23 N, $K_3$=4381.44 N,
$K_4$=1749.54 N, $K_5$=56860.1 N, $K_6$=58318.1 N.

For simplicity, for all bars in the element, the elastic moduli of all bar materials are selected to be the same value, i.e., $E$=100000 N/mm$^2$, then the cross-sectional areas of the bars are:
$A_1$=$K_1/E$ =0.0538144 mm$^2$, $A_2$=$K_2/E$ =0.0247423 mm$^2$, $A_3$=$K_3/E$ =0.0438144 mm$^2$,
$A_4$=$K_4/E$ =0.0174954 mm$^2$, $A_5$=$K_5/E$ =0.568601 mm$^2$, $A_6$=$K_6/E$ =0.583181 mm$^2$.
where, $A_1$ is the cross-sectional area of one parallel bar parallel to the 1-direction,
$A_2$ is the cross-sectional area of one parallel bar parallel to the 2-direction,
$A_3$ is the cross-sectional area of one parallel bar parallel to the 3-direction,
$A_4$ is the cross-sectional area of one oblique bar parallel to the 1-2 plane,
$A_5$ is the cross-sectional area of one oblique bar parallel to the 3-1 plane,
$A_6$ is the cross-sectional area of one oblique bar parallel to the 2-3 plane.

The cross-sectional area of each bar is not negative, so there is no need to adjust the dimensions of the element or modify the design values of the elastic constants of this metamaterial. For simplicity, the cross-sectional shapes of all bars are selected to be circular. Obviously, it is easy to determine the cross-sectional size of each bar. Select the first structure from the two structures of the element. Without drawing, after simple estimation, it is easy to determine that within the space range of one element, the cross-sectional size of each bar is appropriate, and each bar can be connected to form one element. In fact, if the elastic modulus of each bar material in the element is selected to be large enough, the cross-sectional area and cross-sectional size of each bar will be small enough, then within the space range of one element, each bar can always be connected to form one element. The elements are arrayed periodically in space, and the adjacent elements share their adjacent nodes, thus forming this metamaterial. The connection mode between the bars in the element is the hinged connection.

### 2.2 Example 2

The connection mode between the bars in the element is the rigid connection, and the others are exactly the same as in Example 1.

### 2.3 Example 3

Compared with Example 2, Example 3 remains unchanged except that the structure of the element is changed from the first structure to the second structure.

### 2.4 Example 4

For the design values of the 9 elastic constants of this metamaterial, it is known that $v_{12}=0$, $E_3=1000$ N/mm$^2$, then freely select 4 independent elastic constants, $v_{23}=0.4$, $v_{13}=0.3$, $E_1=800$ N/mm$^2$, $E_2=1200$ N/mm$^2$, and the other 3 elastic constants are obtained from Formula (1), i.e., $G_{12}=156.522$ N/mm$^2$, $G_{31}=326.087$ N/mm$^2$, $G_{23}=652.174$ N/mm$^2$. Select the dimensions in the three directions of the element: $l_1=10$ mm, $l_2=11$ mm, $l_3=12$ mm, then calculate the axial stiffness of each bar in the element. For all bars in the element, the elastic moduli of all bar materials are selected as $E=50000$ N/mm$^2$, the cross-sectional areas of the bars are: $A_1=0.344820$ mm$^2$, $A_2=0.465387$ mm$^2$, $A_3=0.0621443$ mm$^2$, $A_4=0.280493$ mm$^2$, $A_5=0.569639$ mm$^2$, $A_6=1.06568$ mm$^2$. Select that the cross-sectional shapes of all bars are square. Select the first structure from the two structures of the element. The connection mode between the bars in the element is the hinged connection, obviously, any bar in the element can be replaced by a spring or a variable cross-section bar according to the principle of keeping the length and stiffness coefficient of the bar unchanged.

### 2.5 Example 5

For the design values of the 9 elastic constants of this metamaterial, it is known that $v_{12}=0$, $v_{23}=0.5$, $v_{13}=0.5$, $E_1=100$ N/mm$^2$, $E_2=100$ N/mm$^2$, $E_3=100$ N/mm$^2$, then the other 3 elastic constants are obtained from Formula (1), i.e., $G_{12}=50$ N/mm$^2$, $G_{31}=100$ N/mm$^2$, $G_{23}=100$ N/mm$^2$. Select the dimensions in the three directions of the element: $l_1=l_2=l_3=4$ mm, then calculate the axial stiffness of each bar in the element. For all bars in the element, the elastic moduli of all bar materials are selected as $E=10000$ N/mm$^2$, the cross-sectional areas of the bars are: $A_1=A_2=A_3=0$ mm$^2$, $A_4=0.0565685$ mm$^2$, $A_5=0.113137$ mm$^2$, $A_6=0.113137$ mm$^2$. Select that the cross-sectional shapes of all bars are circular. Select the second structure from the two structures of the element. The connection mode between the bars in the element is the rigid connection.

### 2.6 Example 6

For the design values of the 9 elastic constants of this metamaterial, it is known that $v_{12}=v_{23}=v_{13}=0$, $E_1=150$ N/mm$^2$, $E_2=100$ N/mm$^2$, $E_3=200$ N/mm$^2$, then the other 3 elastic constants are obtained from Formula (1), i.e., $G_{12}=G_{31}=G_{23}=0$ N/mm$^2$. Select the dimensions in the three directions of the element: $l_1=l_2=l_3=10$ mm, then calculate the axial stiffness of each bar in the element. For all bars in the element, the elastic moduli of all bar materials are selected as $E=10000$ N/mm$^2$, the cross-sectional areas of the bars are: $A_1=0.375$ mm$^2$, $A_2=0.25$ mm$^2$, $A_3=0.5$ mm$^2$, $A_4=A_5=A_6=0$ mm$^2$. Select that the cross-sectional shapes of all bars are circular. Select the first structure from the two structures of the element. The connection mode between the bars in the element is the rigid connection.

### 2.7 Actual values of elastic constants of the metamaterial

In various textbooks on elasticity or mechanics of composite materials, there are 6 Poisson's ratios for orthotropic materials, and there are two completely opposite definitions of Poisson's ratio, one of which is used in this paper, and the corresponding generalized Hooke's law is shown in Formula (4).

$$\begin{bmatrix} \varepsilon_1 \\ \varepsilon_2 \\ \varepsilon_3 \\ \gamma_{23} \\ \gamma_{31} \\ \gamma_{12} \end{bmatrix} = \begin{bmatrix} \frac{1}{E_1} & \frac{-v_{12}}{E_2} & \frac{-v_{13}}{E_3} & 0 & 0 & 0 \\ \frac{-v_{21}}{E_1} & \frac{1}{E_2} & \frac{-v_{23}}{E_3} & 0 & 0 & 0 \\ \frac{-v_{31}}{E_1} & \frac{-v_{32}}{E_2} & \frac{1}{E_3} & 0 & 0 & 0 \\ 0 & 0 & 0 & \frac{1}{G_{23}} & 0 & 0 \\ 0 & 0 & 0 & 0 & \frac{1}{G_{31}} & 0 \\ 0 & 0 & 0 & 0 & 0 & \frac{1}{G_{12}} \end{bmatrix} \begin{bmatrix} \sigma_1 \\ \sigma_2 \\ \sigma_3 \\ \tau_{23} \\ \tau_{31} \\ \tau_{12} \end{bmatrix} \quad (4)$$

Three Poisson's ratios can be represented by another three Poisson's ratios and three elastic moduli, i.e., $v_{21}=v_{12}E_1/E_2$, $v_{31}=v_{13}E_1/E_3$, $v_{32}=v_{23}E_2/E_3$.

Using the finite element method, the actual values of the elastic constants of the metamaterials in Examples 1 to 6 are determined. The following will illustrate, taking one element in each example, and calculating it under uniaxial tension in the 2-direction and pure shear in the 2-3 plane, respectively.

For the example 1, the computational model of the element under uniaxial tension in the 2-direction is shown in Fig. 2, and the computational model of the element under pure shear in the 2-3 plane is shown in Fig. 3; where, the load of the element is determined according to the following method, if the element is subjected to uniaxial tension in the 2-direction, the normal stress $\sigma_2=40$ N/mm² is arbitrarily taken, then when using the finite element analysis, the concentrated force applied to each node $F=\sigma_2\times(l_1\times l_3)\div 4=40\times(10\times 10)\div 4=1000$ N; if the element is subjected to pure shear in the 2-3 plane, the shear stress $\tau_{23}=\tau_{32}=80$ N/mm² is arbitrarily taken, then when using the finite element analysis, the concentrated force applied to each node $F_1=\tau_{23}\times(l_1\times l_3)\div 4=80\times(10\times 10)\div 4=2000$ N, $F_2=\tau_{32}\times(l_1\times l_2)\div 4=80\times(10\times 10)\div 4=2000$ N.

For the examples 2 to 6, according to the same method as the example 1, determine the loads of the element and the computational model of the element. For the element in the example 2 or the example 3, the loads under uniaxial tension in the 2-direction and pure shear in the 2-3 plane, respectively, are exactly the same as those in the example 1. For the example 4, take the normal stress $\sigma_2=120$ N/mm² under uniaxial tension in the 2-direction, and take the shear stress $\tau_{23}=\tau_{32}=88$ N/mm² under pure shear in the 2-3 plane. For the example 5, take the normal stress $\sigma_2=10$ N/mm² under uniaxial tension in the 2-direction, and take the shear stress $\tau_{23}=\tau_{32}=10$ N/mm² under pure shear in the 2-3 plane. For the example 6, take the normal stress $\sigma_2=10$ N/mm² under uniaxial tension in the 2-direction, and take the shear stress $\tau_{23}=\tau_{32}=0.1$ N/mm² under pure shear in the 2-3 plane.

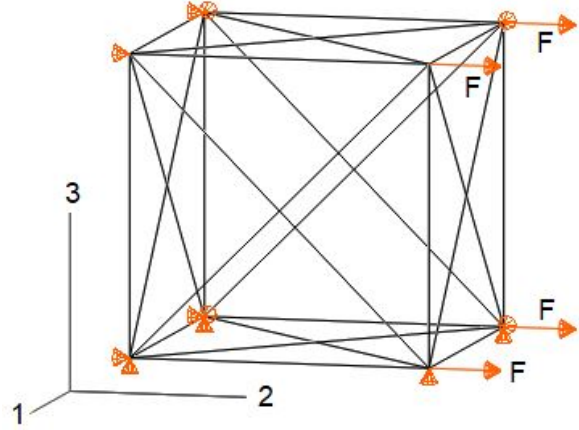

Fig. 2 The computational model of the element in Example 1 under uniaxial tension in the 2-direction

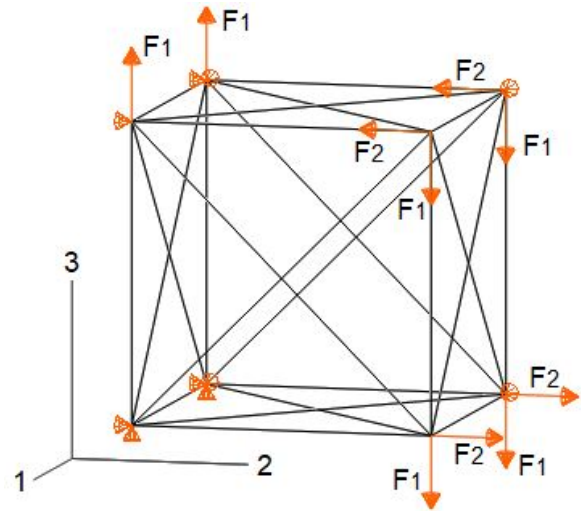

Fig. 3 The computational model of the element in Example 1 under pure shear in the 2-3 plane

Using the finite element software, the strains of the elements in Examples 1 to 6 under uniaxial tension in the 2-direction and pure shear in the 2-3 plane were obtained respectively, as shown in Table 1. The strain in each principal material direction of the element can be calculated from the nodal displacements, or obtained from the axial strain of the bar parallel to the principal material direction, and the shear strain of the element can be calculated from the nodal displacements of the element. In addition, for the metamaterial in each example, one corresponding solid element can also be taken for comparative analysis, that is, the elastic constants of this solid element, take the design values of the elastic constants given in this example; the dimensions, loads and boundary conditions of this solid element and the metamaterial element in this example are exactly the same; or the strain is calculated from the stress based on the above generalized Hooke's law; it can be found that the strain of the solid element is the same or close to the corresponding strain in Table 1.

The design and actual values of the elastic constants of the metamaterials in Examples 1 to 6 are shown in Table 2. Where, the design values in Table 2 are the design values of the elastic constants of the metamaterial in each example, and the actual values are calculated according to the definition of the elastic constants and Table 1. Taking Example 1 to illustrate, the actual values of the elastic constants are: $v_{12}=-\varepsilon_1/\varepsilon_2=-0.05/0.1=-0.5$, $v_{32}=-\varepsilon_3/\varepsilon_2=0.068/0.1=0.68$, $E_2=\sigma_2/\varepsilon_2=40/0.1=400$, $G_{23}=|\tau_{23}/\gamma_{23}|=80/0.0970=824.742$.

Table 1　The strains of the elements in Examples 1 to 6 under uniaxial tension in the 2-direction and pure shear in the 2-3 plane, respectively

| strain / Example | under uniaxial tension in the 2-direction | | | under pure shear in the 2-3 plane |
|---|---|---|---|---|
| | strain $\varepsilon_1$ in the 1-direction | strain $\varepsilon_2$ in the 2-direction | strain $\varepsilon_3$ in the 3-direction | shear strain $\gamma_{23}$ |
| Example 1 | 0.05 | 0.1 | -0.068 | -0.0970 |
| Example 2 | 0.0495 | 0.0994 | -0.0674 | -0.0969 |
| Example 3 | 0.0465 | 0.0959 | -0.0640 | -0.0969 |
| Example 4 | 0 | 0.1 | -0.048 | -0.1349 |
| Example 5 | -0.00062 | 0.0980 | -0.0481 | -0.0998 |
| Example 6 | 0 | 0.1 | 0 | -0.3273 |

Table 2　The design and actual values of the elastic constants of the metamaterials in Examples 1 to 6

| elastic constant / Example | under uniaxial tension in the 2-direction | | | | | | under pure shear in the 2-3 plane | |
|---|---|---|---|---|---|---|---|---|
| | $v_{12}$ | | $v_{32}$ | | $E_2$ (N/mm$^2$) | | $G_{23}$ (N/mm$^2$) | |
| | design value | actual value | design value | actual value | design value | actual value | design value | actual value |
| Example 1 | -0.5 | -0.5 | 0.68 | 0.68 | 400 | 400 | 824.7 | 824.7 |
| Example 2 | -0.5 | -0.498 | 0.68 | 0.678 | 400 | 402 | 824.7 | 825.6 |
| Example 3 | -0.5 | -0.485 | 0.68 | 0.667 | 400 | 417 | 824.7 | 825.6 |
| Example 4 | 0 | 0 | 0.48 | 0.48 | 1200 | 1200 | 652.2 | 652.2 |
| Example 5 | 0 | 0.006 | 0.5 | 0.491 | 100 | 102 | 100 | 100.2 |
| Example 6 | 0 | 0 | 0 | 0 | 100 | 100 | 0 | 0.31 |

　　It can be seen from Table 2 that if the connection mode between the bars in the element is hinged connection and the structure of the element is the first structure, then the actual value of the elastic constant of the metamaterial is exactly the same as the design value, e.g., Examples 1 and 4; if the connection mode between the bars in the element is rigid connection, then whether the first structure or the second structure is adopted for the structure of the element, there is a deviation between the actual value of the elastic constant of the metamaterial and the design value, but the design accuracy is good, e.g., Example 2, Example 3, Example 5, Example 6. In general, for the manufactured metamaterial, under the premise of ensuring the stability of the element, if the actual connection mode between the bars in the element is closer to the hinged connection, the design accuracy will be higher.

　　For the example 1, the deformation of the element under uniaxial tension in the 2-direction is shown in Fig. 4, and the deformation of the element under pure shear in the 2-3 plane is shown in Fig. 5, which are consistent with the theoretical predictions. Fig. 6, Fig. 7 and Fig. 8 show the elements of the metamaterials in Example 3, Example 5, and Example 6 respectively, the element structure of the metamaterial in each example can be seen intuitively, and it is obvious that it can be easily manufactured. Generally, the structure of the element adopts the second structure, which has better convenience for the manufacture of metamaterials.

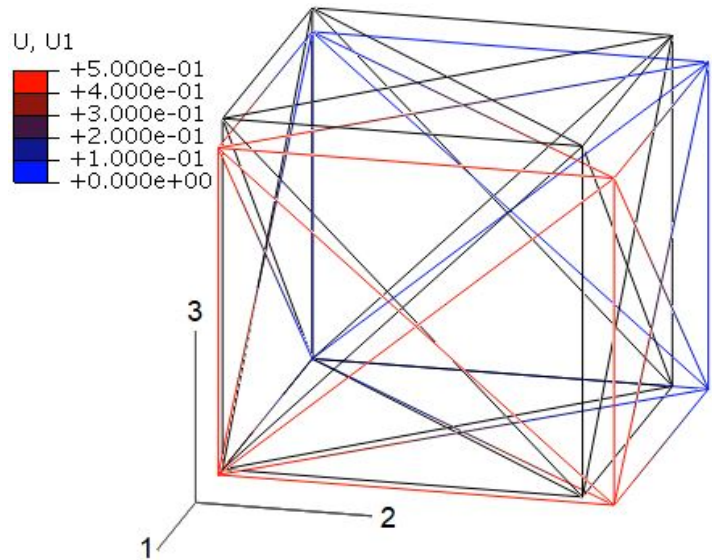

Fig. 4 The deformation of the element in Example 1 under uniaxial tension in the 2-direction

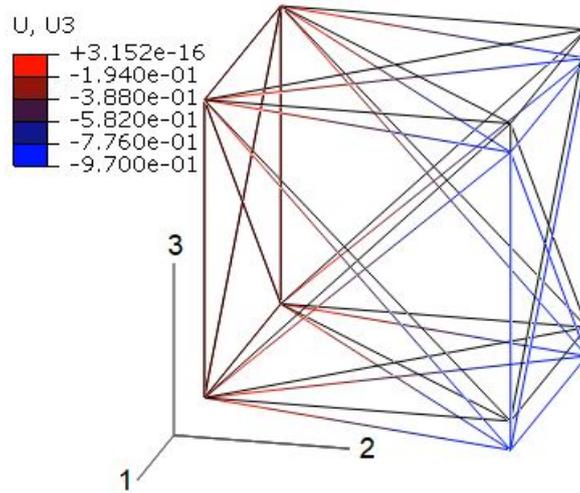

Fig. 5 The deformation of the element in Example 1 under pure shear in the 2-3 plane

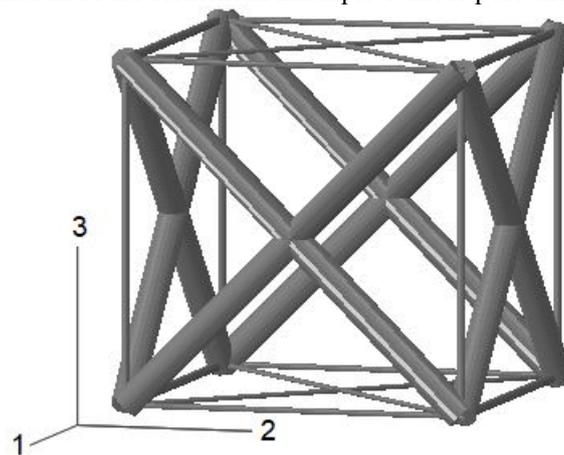

Fig. 6 The element of the metamaterial in Example 3

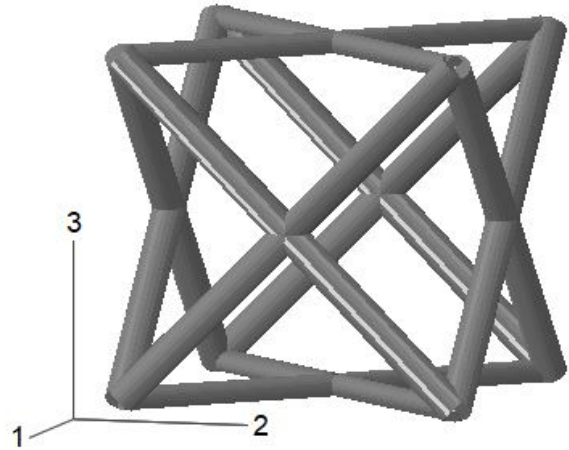

Fig. 7 The element of the metamaterial in Example 5

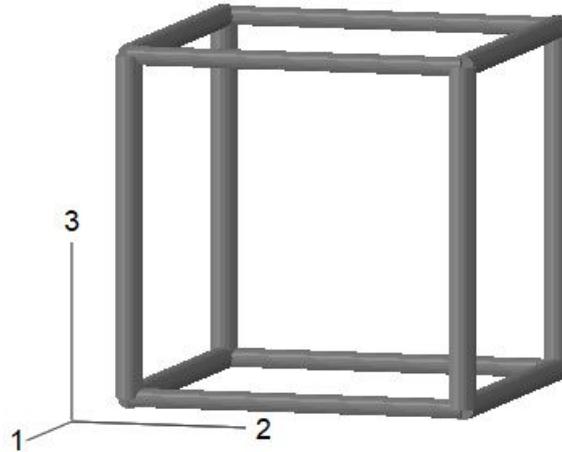

Fig. 8 The element of the metamaterial in Example 6

**3 Expansion of metamaterials**

The space truss element in Reference 17 is applied to materials with 9 independent elastic constants, while the space truss element in Reference 16 is a special case of the space truss element in Reference 17. Using the same method as above, an orthotropic metamaterial can be proposed based on the space truss element model in Reference 17, but the structure of the element is complex, which is inconvenient to manufacture. In addition, using the same method as above, a planar metamaterial can be proposed based on the plane stress element model in Reference 16 or 17.

**4 Conclusion**

The orthotropic metamaterial proposed in this paper has 6 independent elastic constants, so there is a large degree of freedom in metamaterial design. Using a simple design method, a metamaterial with tailored elastic constants can be designed. Therefore, it has great application value in the fields of mechanical metamaterials, elastic wave metamaterials, acoustic metamaterials, and seismic metamaterials, and has also laid the foundation for realizing the dream of controlling elastic waves, acoustic waves and vibrations. The element structure of the metamaterial is very simple and easy to manufacture. In addition, using the same method, it can be extended to 3D metamaterials with 9 independent elastic constants and planar metamaterials.